\documentclass[prb,twocolumn,nofootinbib]{revtex4-2}
\usepackage{amsmath}
\usepackage{amssymb}
\usepackage{graphicx}
\usepackage{hyperref}
\usepackage{newtxtext}
\usepackage[varvw]{newtxmath}
\usepackage{bbm}
\usepackage{bbold}
\usepackage{tabularx}
\usepackage{comment}
\usepackage[position=top,caption=false]{subfig} % loads caption3
\usepackage{ragged2e} % for the \justifying macro

\hypersetup{
    colorlinks,
    linkcolor={blue},
    citecolor={blue},
    urlcolor={blue}
}

\graphicspath{{./}{./figs/}}

\newcommand{\rmi}{\mathrm{i}}
\newcommand{\rmd}{\mathrm{d}}

\newcommand{\Nf}{N_\mathrm{f}}

\DeclareMathOperator{\Tr}{Tr}

\DeclareMathAlphabet{\mymathbb}{U}{BOONDOX-ds}{m}{n}

\DeclareCaptionJustification{justified}{\justifying}

\usepackage[dvipsnames]{xcolor}
\usepackage[normalem]{ulem}

\begin{document}
\title{Gross-Neveu-Heisenberg criticality from \(\boldsymbol{2+\epsilon}\) expansion}
%arXiv:Gross-Neveu-Heisenberg criticality from $2+\epsilon$ expansion

\author{Konstantinos Ladovrechis}

\author{Shouryya Ray}

\author{Tobias Meng}

\author{Lukas Janssen}

\affiliation{Institut f\"ur Theoretische Physik and W\"urzburg-Dresden Cluster of Excellence ct.qmat, TU Dresden, 01062 Dresden, Germany}

%%%%%%%%%%%%%%%%%%%%%%%%%%%%%%%%%%%%%%%%%%%%%%%%%%%%%%%%%%%%%%%%%%%%%%%
\begin{abstract}
The Gross-Neveu-Heisenberg universality class describes a continuous quantum phase transition between a Dirac semimetal and an antiferromagnetic insulator. Such quantum critical points have originally been discussed in the context of Hubbard models on $\pi$-flux and honeycomb lattices, but more recently also in Bernal-stacked bilayer models, of potential relevance for bilayer graphene.
Here, we demonstrate how the critical behavior of this fermionic universality class can be computed within an $\epsilon$ expansion around the lower critical space-time dimension of two. This approach is complementary to the previously studied expansion around the upper critical dimension of four. The crucial technical novelty near the lower critical dimension is the presence of different four-fermion interaction channels at the critical point, which we take into account in a Fierz-complete way.
By interpolating between the lower and upper critical dimensions, we obtain improved estimates for the critical exponents in 2+1 space-time dimensions.
For the situation relevant to single-layer graphene, we find an unusually small leading-correction-to-scaling exponent, arising from the competition between different interaction channels.
This suggests that corrections to scaling may need to be taken into account when comparing analytical estimates with numerical data from finite-size extrapolations.
\end{abstract}
%%%%%%%%%%%%%%%%%%%%%%%%%%%%%%%%%%%%%%%%%%%%%%%%%%%%%%%%%%%%%%%%%%%%%%%

\date{\today}

\maketitle

%%%%%%%%%%%%%%%%%%%%%%%%%%%%%%%%%%%%%%%%%%%%%%%%%%%%%%%%%%%%%%%%%%%%%%%
\section{Introduction}
\label{sec:introduction}
%%%%%%%%%%%%%%%%%%%%%%%%%%%%%%%%%%%%%%%%%%%%%%%%%%%%%%%%%%%%%%%%%%%%%%%

Fermionic quantum critical points are continuous quantum phase transitions that are driven by interactions between gapless fermionic degrees of freedom. They can be viewed as the simplest examples of quantum critical points that do not exhibit classical analogs.
Such transitions were originally discussed in the context of toy models, mimicking aspects relevant to high-energy physics, such as chiral symmetry breaking and spontaneous mass generation~\cite{gross74}, non-perturbative renormalizability~\cite{gawedzki85, rosenstein89, zinnjustin91}, and asymptotic safety~\cite{braun11}.
The poster child of fermionic quantum criticality is embodied by the $(2+1)$-dimensional Gross-Neveu-Ising transition, across which massless Dirac fermions in two spatial dimensions acquire an interaction-induced mass gap as a consequence of a spontaneous $\mathbb Z_2$ symmetry breaking~\cite{%
hands93, wang14, li15, hesselmann16, huffman17, huffman20,
rosenstein93, zerf17,
rosa01, hoefling02, janssen14, vacca15, knorr16,
gracey16, 
gracey18b,
iliesiu18}.
From a field-theoretical viewpoint, a crucial simplicity of the Gross-Neveu-Ising transition is the absence at criticality of any further four-fermion interaction channel at all orders in perturbation theory~\cite{gehring15}.
This allows one to compute loop corrections to high orders not only in the vicinity of the upper critical space-time dimension of four~\cite{zerf17}, but also near the lower critical space-time dimension of two~\cite{gracey16}.

In many physically relevant lattice realizations of fermion quantum criticality, however, the symmetry that spontaneously breaks across the transition is continuous, and governed by a vector order parameter.
A well-known example is given by the Hubbard model on the honeycomb lattice~\cite{sorella92, herbut06, assaad14}, which realizes as function of on-site repulsion $U$ a direct and continuous transition between a symmetric Dirac semimetal at small $U$ and an antiferromagnetic insulator at large $U$. In the strong-coupling phase, the fermionic spectrum is gapped out and the SU(2) spin symmetry is spontaneously broken.
The transition is expected to fall into the Gross-Neveu-Heisenberg universality class~\cite{herbut09}, which has been heavily discussed in recent years~\cite{%
toldin15, liu19, liu21,
otsuka16, otsuka20, 
xu21,
buividovich18, buividovich19, ostmeyer20, ostmeyer21, 
lang19,
janssen14, knorr18,
zerf17,
gracey18a}.
The transition between nematic and coexistent nematic-antiferromagnetic orders on the Bernal-stacked honeycomb bilayer has recently been identified as another potential realization of Gross-Neveu-Heisenberg criticality, albeit with the number of fermion degrees of freedom doubled in comparison with the single-layer case~\cite{ray21b}.
Another possible realization on the Bernal-stacked honeycomb bilayer has been proposed for the transition between the trigonal-warping-induced Dirac semimetal and the antiferromagnetic insulator~\cite{pujari16, ray18}. In this latter case, each quadratic band touching point present in the noninteracting limit for vanishing trigonal warping splits into four Dirac cones, leading to a quadrupled number of fermion degrees of freedom in comparison with the the single-layer case~\cite{ray18}.

Similar to classical universality classes, each fermionic quantum universality class is characterized by a unique set of universal critical exponents.
For the relativistic Gross-Neveu-type criticalities, the dynamical critical exponent $z=1$, leaving three independent quantities $1/\nu$, $\eta_\phi$, and $\eta_\psi$, corresponding to the correlation-length exponent and the order-parameter and fermion anomalous dimensions.
While for the Gross-Neveu-Ising case provisional convergence of the predictions of the different methods appears within reach~\cite{ihrig18, huffman20}, the disagreement among the different literature results in case of the Gross-Neveu-Heisenberg criticality remains significant, see, e.g., Ref.~\cite{liu21} for a recent overview.
On the field-theoretical side, a major challenge is the fact that the perturbative series are at best asymptotically convergent, requiring appropriate resummation schemes. 
In the Gross-Neveu-Ising case, a significant step forward has been the possibility to employ interpolational schemes that make use of the known expansions near the lower and upper critical dimensions simultaneously~\cite{janssen14, ihrig18}. 
In the Gross-Neveu-Heisenberg case, however, different four-fermion interaction channels are generically present at criticality, necessitating a Fierz-complete study that deals with these channels in an unbiased way~\cite{jaeckel03}.

In this work, we provide such an analysis.
We study the renormalization group (RG) fixed-point structure of the theory space defined by the symmetries of the Gross-Neveu-Heisenberg field theory within an expansion around the lower critical space-time dimension of two.
After Fierz reduction, this space is spanned by six independent four-fermion couplings. We identify the fixed point corresponding to Gross-Neveu-Heisenberg criticality and determine the corresponding quantum critical behavior in terms of the correlation-length exponent $1/\nu$ and the order-parameter anomalous dimensions $\eta_\phi$ to one-loop order.
The fermion anomalous dimension $\eta_\psi$ is computed to two-loop order.
To arrive at these results, we derive general formulae for the order-parameter and fermion anomalous dimensions that we expect to be of use for the community also beyond the particular theory space studied in this work.
Our results near the lower critical dimension allow us to provide improved estimates for the exponents in the physical situation in 2+1 space-time dimensions by employing an interpolational resummation scheme that takes also previous results near the upper critical dimension~\cite{zerf17} into account.

The remainder of this paper is organized as follows: In the next section, we determine the theory space of the Gross-Neveu-Heisenberg model, its symmetries, and a corresponding Fierz-complete basis.
The RG flow and the fixed-point structure are discussed in Sec.~\ref{sec:renormalization}.
In Sec.~\ref{sec:exponents}, we derive the general formulae to compute the order-parameter and fermion anomalous dimensions $\eta_\phi$ and $\eta_\psi$ to one-loop and two-loop order, respectively, and use these to provide a complete set of critical exponents near the lower critical dimension.
Estimates for the exponents in 2+1 space-time dimensions using an interpolational resummation scheme are presented in Sec.~\ref{sec:estimates}.
Section~\ref{sec:conclusions} contains our conclusions and outlook.

%%%%%%%%%%%%%%%%%%%%%%%%%%%%%%%%%%%%%%%%%%%%%%%%%%%%%%%%%%%%%%%%%%%%%%%
\section{Gross-Neveu-Heisenberg theory space}
\label{sec:gross-neveu-heisenberg}
%%%%%%%%%%%%%%%%%%%%%%%%%%%%%%%%%%%%%%%%%%%%%%%%%%%%%%%%%%%%%%%%%%%%%%%

%%%%%%%%%%%%%%%%%%%%%%%%%%%%%%%%%%%%%%%%%%%%%%%%%%%%%%%%%%%%%%%%%%%%%%%
\subsection{Microscopic model}
\label{subsec:model}
%%%%%%%%%%%%%%%%%%%%%%%%%%%%%%%%%%%%%%%%%%%%%%%%%%%%%%%%%%%%%%%%%%%%%%%

Within a purely fermionic formulation, the Gross-Neveu-Heisenberg model can be defined via the Euclidean microscopic action~\cite{rosenstein93, herbut09, janssen14, gracey18a, ray21b}
\begin{align}\label{eq:GNH}
    S_{\text{GNH}} & =\int\! \rmd^D{x}\Bigl\{\bar{\psi}^{\alpha}(\gamma_\mu\otimes\mathbb{1}_2)\partial_\mu\psi^{\alpha}
    \nonumber\\ &\quad    
    -\frac{{g}_1}{2\Nf}\left[\bar{\psi}^{\alpha}(\mathbb{1}_2\otimes\vec\sigma)\psi^{\alpha}\right]^2\Bigr\}
\end{align}
in $D=2+\epsilon$ space-time dimensions.
In the above equation, the space-time index $\mu=0,\dots, D - 1$, the flavor index $\alpha=1, \dots ,\Nf$, with $\Nf$ counting the number of four-component Dirac fermions, and $\vec\sigma = (\sigma_x, \sigma_y, \sigma_z)$ denotes the vector of $2\times 2$ Pauli matrices.
Here and in what follows, if not stated otherwise, summation over repeated indices is implicitly assumed. 
In two dimensions, we employ an irreducible two-dimensional representation of the Clifford algebra $\{\gamma_\mu,\gamma_\nu\}=2\delta_{\mu\nu}\mathbb{1}_2$, such as
\begin{equation}
\label{eq:gamma}
    \gamma_0=\begin{pmatrix} 0 & -\rmi \\
                              \rmi & 0
              \end{pmatrix}
    \qquad \text{and} \qquad
    \gamma_1=\begin{pmatrix} 0 & 1 \\
                              1 & 0
              \end{pmatrix}.
\end{equation}
The Dirac conjugate field is defined as $\bar{\psi}\equiv \psi^{\dagger}(\gamma_0\otimes\mathbb{1}_2)$.
The Gross-Neveu-Heisenberg coupling $g_1$ has mass dimension $[g_1] = 2-D$. The coupling becomes dimensionless, and the theory perturbatively renormalizable, at $D = 2$. Thus, $D=2$ defines a critical space-time dimension that can further be identified as the lower critical dimension, around which we expand. 
This approach is complementary to the previously studied expansion around the upper critical dimension of four~\cite{rosenstein93, herbut09, zerf17}.

Within the large-$\Nf$ expansion, an ultraviolet completion of the above model exists for all dimensions $2<D<4$~\cite{gracey18a}, and we assume this property to hold also for finite $\Nf$.
A potential complication is the possibility of emergent evanescent operators that may in principle be generated within the perturbative expansion~\cite{bennett99, ali01, gracey16}. 
At the one-loop level, these induce shifts in the $\beta$ functions, which cancel with corresponding contributions from the two-loop diagrams~\cite{bondi90}. 
As for the critical exponents, we therefore expect the emergence of evanescent operators to play a role only beyond the leading-order estimates computed in the present work.

In the (2+1)-dimensional realization of the model on the single-layer honeycomb lattice~\cite{herbut06, herbut09}, the number of four-component fermion flavors is $\Nf=2$. The criticality between nematic and coexistent nematic-antiferromagnetic orders in bilayer graphene corresponds to $\Nf=4$~\cite{ray21b}. At the transition between the warping-induced spin-$1/2$ Dirac semimetal and the antiferromagnetic insulator on the honeycomb bilayer, the number of gapless four-component flavors is $\Nf=8$~\cite{ray18}.

%%%%%%%%%%%%%%%%%%%%%%%%%%%%%%%%%%%%%%%%%%%%%%%%%%%%%%%%%%%%%%%%%%%%%%%
\subsection{Symmetries}
\label{subsec:symmetries}
%%%%%%%%%%%%%%%%%%%%%%%%%%%%%%%%%%%%%%%%%%%%%%%%%%%%%%%%%%%%%%%%%%%%%%%

In contrast to the Gross-Neveu-Ising case~\cite{gracey16}, the microscopic action defined in Eq.~\eqref{eq:GNH} is not closed under RG transformations.
Already at the one-loop order, fluctuations induce new interaction channels that need to be taken into account in a consistent way.
The possible newly generated terms are, however, strongly constrained by the symmetries of the microscopic action.
These symmetries are:

\paragraph{Lorentz invariance:}
In two Euclidean space-time dimensions, the Dirac spinors transform as
\begin{equation}
    \psi(x)\mapsto e^{-\rmi \varepsilon \frac{\rmi}{4}[\gamma_0,\gamma_1]\otimes\mathbb{1}_2}\psi(x'),
\end{equation}
where we have suppressed the flavor index for simplicity. The space-time coordinate $x = (x^\mu)$ transforms as $x^\mu \mapsto {x'}^{\mu}=\Lambda^{\mu}{}_{\nu}x^{\nu}$, with rotation matrix $(\Lambda^\mu{}_\nu) \in \mathrm{O}(2)$.

\paragraph{Flavor symmetry:}
\begin{equation}
    \psi^{\alpha}\mapsto U^{\alpha\beta}\psi^{\beta},
%    \qquad\bar{\psi}^{\alpha}\mapsto\bar{\psi}^{\beta}(U^{\alpha\beta})^{\dag},
\end{equation}
with unitary matrix $U \in \operatorname{U}(\Nf)$.

\paragraph{$\mathbbm Z_2$ chiral symmetry:}
\begin{equation}
    \psi \mapsto (\gamma_5\otimes\mathbb{1}_2)\psi, 
%    \qquad\bar{\psi}\mapsto -\bar{\psi}(\gamma_5\otimes\mathbb{1}_2),
\end{equation}
where $\gamma_5 = \rmi \gamma_0 \gamma_1$ denotes the chiral matrix, which is diagonal in the representation of Eq.~\eqref{eq:gamma}.
Note that $\bar\psi \mapsto - \bar\psi (\gamma_5 \otimes \mathbb 1_2)$, such that the mass bilinears $\bar\psi\psi$ and $\rmi\bar\psi(\gamma_5 \otimes \mathbb 1_2)\psi$ are odd under chiral symmetry.

\paragraph{SU(2) spin symmetry:}
\begin{equation} \label{eq:spinSUtwo}
    \psi^{\alpha}\mapsto e^{\rmi\phi\vec{n}\cdot(\mathbb{1}_2\otimes\vec{\sigma})}\psi^{\alpha},
%    \qquad\bar{\psi}^{\alpha}\mapsto \bar{\psi}^{\alpha}e^{-\rmi\phi\vec{n}(\mathbb{1}_2\otimes\vec{\sigma})},
\end{equation}
under which the Heisenberg bilinear $\bar{\psi}(\mathbb{1}_2\otimes\vec\sigma)\psi$ transforms as a vector.

\paragraph{Time-reversal symmetry:}
\begin{equation}
    \psi(x) \mapsto \mathcal T \psi(x),
\end{equation}
with the time-reversal operator $\mathcal{T}=(\gamma_1\otimes\sigma_y)\mathcal{K}$ in Euclidean time, where $\mathcal{K}$ denotes complex conjugation.
We note that the Heisenberg bilinear $\bar{\psi}(\mathbb{1}_2\otimes\vec\sigma)\psi$ is odd under time reversal, in agreement with its lattice realizations in 2+1 dimensions, where it corresponds to antiferromagnetic orders~\cite{herbut06, herbut09, ray21b}.
The mass bilinears $\bar\psi\psi$ and $\rmi\bar\psi(\gamma_5\otimes \mathbb 1_2)\psi$ are even under time reversal.

\paragraph{Inversion symmetry:}
\begin{equation}
    \psi(x) \mapsto \mathcal{I} \psi(x'),
\end{equation}
with the inversion operator $\mathcal{I}=\gamma_0\otimes\mathbb{1}_2$, and the space-time coordinate $x = (x^0, x^1)$ transforming as $x \mapsto x' = (x^0, -x^1)$.
While the standard mass term $\bar\psi\psi$ is inversion symmetric, the bilinear $\rmi\bar\psi\gamma_5\psi$ is odd under inversion.

%%%%%%%%%%%%%%%%%%%%%%%%%%%%%%%%%%%%%%%%%%%%%%%%%%%%%%%%%%%%%%%%%%%%%%%
\subsection{Classification of four-fermion operators}
\label{subsec:classification}
%%%%%%%%%%%%%%%%%%%%%%%%%%%%%%%%%%%%%%%%%%%%%%%%%%%%%%%%%%%%%%%%%%%%%%%

The above symmetries forbid mass terms or other bilinears in the effective action obtained by integrating out fermionic fluctuations.
However, in addition to the Gross-Neveu-Heisenberg four-fermion interaction already present in the microscopic model, Eq.~\eqref{eq:GNH}, there exist other four-fermion terms that feature the same symmetries and thus will generically be generated under the RG.
In order to identify a basis of the theory space, we now classify all possible four-fermion terms according to their symmetries.

Flavor symmetry allows for two different types of four-fermion terms: those having singlet flavor structure $(\bar{\psi}^{\alpha}\mathcal{O}\psi^{\alpha}) (\bar{\psi}^{\beta}\mathcal{Q}\psi^{\beta})$ and those with non-singlet flavor structure $(\bar{\psi}^{\alpha}\mathcal{O}\psi^{\beta}) (\bar{\psi}^{\beta}\mathcal{Q}\psi^{\alpha})$.
Here, $\mathcal{O}$ and $\mathcal{Q}$ denote $4\times 4$ matrices that act on the spinor indices of $\psi$ and $\bar\psi$.
The singlet and non-singlet terms are related to each other via Fierz identities~\cite{gies10, gehring15}, and it is therefore always possible to write the latter as a linear combination of the former. 
Similarly, terms of the form $(\psi^{\alpha \top} \mathcal O \psi^\beta) (\bar\psi^\beta \mathcal Q \bar\psi^{\alpha \top})$ are related to the above terms via Fierz identities, and do not lead to any new independent interaction channels.
For our purposes, it thus suffices to determine the invariant four-fermion terms with flavor singlet structure.
These can be constructed from the symmetry transformation properties of the bilinears
\begin{equation}\label{eq:bilinear}
    \bar{\psi}^{\alpha}\mathcal{O}\psi^{\alpha},
\end{equation}
with $\mathcal{O}$ being a $4\times4$ matrix acting on the spinor indices of $\psi$ and $\bar\psi$. A basis in the sixteen-dimensional space of $4\times4$ matrices is given by the direct product of the basis matrices of the charge and the spin sectors,
\begin{equation}\label{eq:int_basis}
    \{\mathbb{1}_2,\gamma_0,\gamma_1,\gamma_5\}\otimes\{\mathbb{1}_2,\vec\sigma\},
\end{equation}
and any $\mathcal{O}$ can hence be written as a linear combination of these.

We now discuss the transformation properties of these basis matrices.
They can be divided into two groups of eight matrices each, $A=\{\mathbb{1}_2,\gamma_5\}\otimes\{\mathbb{1}_2,\vec\sigma\}$ and $B=\{\gamma_0,\gamma_1\}\otimes\{\mathbb{1}_2,\vec\sigma\}$, which commute and anticommute, respectively, with the chiral matrix $\gamma_5$.
Each group can be further split into sets of spin SU(2) scalars and vectors, respectively, viz., 
$A_\text{S} = (\mathbb 1_2, \gamma_5) \otimes \mathbb 1_2$, 
$A_\text{V} = (\mathbb 1_2, \gamma_5) \otimes \vec \sigma$, 
$B_\text{S} = (\gamma_0, \gamma_1) \otimes \mathbb 1_2$, 
and
$B_\text{V} = (\gamma_0, \gamma_1) \otimes \vec \sigma$.
Any flavor-singlet four-fermion term invariant under both chiral and SU(2) spin symmetries can therefore be written as
$\bar\psi^\alpha \mathcal{O} \psi^\alpha \bar\psi^\beta \mathcal{Q} \psi^\beta$
with $\mathcal{O}$ and $\mathcal{Q}$ from the same set $A_\text{S}$, $A_\text{V}$, $B_\text{S}$, or $B_\text{V}$.
Invariance under time reversal and inversion then implies that $\mathcal{O} = \mathcal{Q}$.
Finally, Lorentz invariance implies that the two different four-fermion terms with $\mathcal{O} = \mathcal{Q} \in B_\text{S}$ appear symmetrically in the Lagrangian with the same coefficients, and equivalently for $\mathcal{O} = \mathcal{Q} \in B_\text{V}$.
Assuming $\Nf > 1$, a Fierz-complete basis of the Gross-Neveu-Heisenberg theory space therefore contains six four-fermion terms,
\begin{align}
    \mathcal L_{\text{int}} & = 
    - \frac{g_1}{2\Nf} \left[\bar{\psi}^{\alpha}(\mathbb{1}_2\otimes\vec\sigma)\psi^{\alpha}\right]^2
    - \frac{g_2}{2\Nf} \left[\bar{\psi}^{\alpha}(\gamma_\mu\otimes\vec\sigma)\psi^{\alpha}\right]^2
    \nonumber \\ & \quad 
    - \frac{g_3}{2\Nf} \left[\bar{\psi}^{\alpha}(\gamma_5\otimes\vec\sigma)\psi^{\alpha}\right]^2
    - \frac{g_4}{2\Nf} \left[\bar{\psi}^{\alpha}(\mathbb{1}_2\otimes\mathbb{1}_2)\psi^{\alpha}\right]^2
    \nonumber \\ & \quad 
    - \frac{g_5}{2\Nf} \left[\bar{\psi}^{\alpha}(\gamma_\mu\otimes\mathbb{1}_2)\psi^{\alpha}\right]^2
    - \frac{g_6}{2\Nf} \left[\bar{\psi}^{\alpha}(\gamma_5\otimes\mathbb{1}_2)\psi^{\alpha}\right]^2,
\label{eq:Lint}
\end{align}
parametrized by the six couplings $\boldsymbol g = (g_1, \dots, g_6)$.%
\footnote{For $\Nf = 1$, there exist further Fierz identities that may reduce the number of indepdenent four-fermion terms.}
Here, $g_1$ corresponds to the Gross-Neveu-Heisenberg coupling~\cite{rosenstein93, herbut09, janssen14, gracey18a, ray21b}, $g_4$ corresponds to the Gross-Neveu-Ising coupling~\cite{rosenstein89, zinnjustin91, hands93, braun11, gehring15, gracey16}, and $g_5$ corresponds to the Thirring coupling~\cite{bondi90, gies10, janssen12}.
In following, we study the flow of the full effective action
\begin{equation}
    S = \int\! \rmd^D x \left[ \bar\psi^\alpha (\gamma_\mu \otimes \mathbb 1_2) \partial_\mu \psi^\alpha + \mathcal L_\text{int} \right],
\end{equation}
out of which $S_\text{GNH}$ is a subspace that we explicitly show to be not closed under the RG.

%%%%%%%%%%%%%%%%%%%%%%%%%%%%%%%%%%%%%%%%%%%%%%%%%%%%%%%%%%%%%%%%%%%%%%%
\section{Renormalization group flow}
\label{sec:renormalization}
%%%%%%%%%%%%%%%%%%%%%%%%%%%%%%%%%%%%%%%%%%%%%%%%%%%%%%%%%%%%%%%%%%%%%%%

%%%%%%%%%%%%%%%%%%%%%%%%%%%%%%%%%%%%%%%%%%%%%%%%%%%%%%%%%%%%%%%%%%%%%%%
\subsection{Flow equations}
%%%%%%%%%%%%%%%%%%%%%%%%%%%%%%%%%%%%%%%%%%%%%%%%%%%%%%%%%%%%%%%%%%%%%%%
%
In order to compute the RG flow equations for the six four-fermion couplings $\boldsymbol{g} = (g_1, \dots, g_6)$, we employ the general one-loop formula derived in \cite{gehring15}.
A straightforward evaluation of the matrix algebra occurring in this formula, using standard computer algebra software, yields the following flow equations, valid for arbitrary $\Nf$,
\begin{align}
    \beta_1 & = \left[\epsilon + \frac{2(2g_2-g_3+g_4+2g_5+g_6)}{\Nf} \right] g_1
    \nonumber \\ & \quad 
    - \frac{2(2\Nf+1)}{\Nf}g^2_1
    + \frac{4(g_3g_5+g_2g_6)}{\Nf}, \label{eq:betafunctions-1} \displaybreak[1] \\
   \beta_2 & = \epsilon g_2
    + \frac{8g^2_2}{\Nf}
    +\frac{2[g_3(g_3+g_4) + g_1(g_1+g_6)]}{\Nf}, \displaybreak[1] \\
   \beta_3 & = \left[\epsilon + \frac{2(g_1 + 2g_2 - g_4 + 2 g_5 - g_6)}{\Nf} \right] g_3
    \nonumber \\ & \quad 
    + \frac{2(2\Nf+1)}{\Nf}g_3^2
    + \frac{4(g_2g_4 + g_1 g_5)}{\Nf}, \displaybreak[1] \\
   \beta_4 & = \left[\epsilon + \frac{2(3g_1+6g_2+3g_3+2g_5+g_6)}{\Nf} \right] g_4
    \nonumber \\ & \quad 
    - \frac{2(2\Nf-1)}{\Nf}g_4^2
    + \frac{4(3g_2g_3+g_5g_6)}{\Nf}, \displaybreak[1] \\
   \beta_5 & = \epsilon g_5+\frac{2(3g_1g_3+g_4g_6)}{\Nf}, \displaybreak[1] \\
   \beta_6 & = \left[\epsilon - \frac{2(3g_1-6g_2+3g_3+g_4-2g_5)}{\Nf} \right] g_6
    \nonumber \\ & \quad 
    + \frac{2(2\Nf-1)}{\Nf}g_6^2
    + \frac{4(3g_1g_2+g_4g_5)}{\Nf}, 
    \label{eq:betafunctions-6}
\end{align}
where $\epsilon = D - 2$ and we have rescaled the couplings as $\ell_1^\text{(F)}g_i/2\pi \mapsto g_i$ for $i=1,\dots,6$, with $\ell_1^\text{(F)}$ a dimensionless regulator-dependent constant. The sign of the $\beta$ functions is defined such that a coupling $g_i$ decreases (increases) in the flow towards the infrared if $\beta_i > 0$ ($\beta_i < 0$).
For $g_1 = g_2 = g_3 = 0$, the above flow equations are consistent with those of Ref.~\cite{bondi90} upon identifying $\ell^\text{(F)}_1 = 1$ for the minimal subtraction scheme employed therein.
We note that the flow equations~\eqref{eq:betafunctions-1}--\eqref{eq:betafunctions-6} are invariant under the exchange of the couplings as $(g_1, g_2, g_3, g_4, g_5, g_6) \leftrightarrow (-g_3, g_2, -g_1, -g_6, g_5, -g_4)$. This property arises from the fact that the chiral matrix $\gamma_5$ anticommutes with the fermion propagator and squares to one.

%%%%%%%%%%%%%%%%%%%%%%%%%%%%%%%%%%%%%%%%%%%%%%%%%%%%%%%%%%%%%%%%%%%%%%%
\subsection{Large-\textit{N}\textsubscript{f} fixed-point structure}
%%%%%%%%%%%%%%%%%%%%%%%%%%%%%%%%%%%%%%%%%%%%%%%%%%%%%%%%%%%%%%%%%%%%%%%
%
The topology of the RG flow is determined by the solutions of the fixed-point equations $\partial_t g_i|_{\boldsymbol{g}^\star}=0$. 
Any nontrivial solution $\boldsymbol{g}^\star \neq 0$ is located at finite distance of order $\mathcal O(\epsilon)$ from the Gaussian fixed point $\boldsymbol{g}^\star = 0$.
While the Gaussian fixed point is fully infrared stable for $\epsilon > 0$, all interacting fixed points have at least one infrared relevant direction.
For initial values of the couplings beyond a certain finite threshold of order $\mathcal O(\epsilon)$, the flow is unstable and diverges at finite RG time, indicating the onset of spontaneous symmetry breaking.
Critical fixed points, which govern the quantum critical behavior of continuous phase transitions, are those with precisely one infrared relevant direction.
Among the different critical fixed points, we are looking for the one corresponding to the onset of spin symmetry breaking via condensation of the SU(2) vector
\begin{equation}
    \langle \bar\psi^\alpha (\mathbbm 1_2 \otimes \vec \sigma) \psi^\alpha \rangle \neq 0.
\end{equation}
This fixed point is readily identified in the large-$\Nf$ limit. In this limit, the individual flow equations for $g_1, g_2, \dots, g_6$ decouple and the fixed-point structure can be computed analytically.
For the set of six quadratic fixed-point equations, there may be at most $2^6 = 64$ possibly degenerate and/or complex solutions. However, in the large-$\Nf$ limit, the quadratic term $\propto g_2^2$ vanishes in the flow equation for $g_2$. The same is true, albeit also beyond the large-$\Nf$ limit, for the quadratic term $\propto g_5^2$ in the flow equation of the Thirring coupling $g_5$~\cite{bondi90}. The absence of such quadratic term can be understood as a fixed point located at infinite coupling $g_2^\star \to \infty$ and $g_5^\star \to \infty$, respectively. In fact, new fixed points located at $g_2^\star \propto \Nf$ emerge upon the inclusion of the $\mathcal O(1/\Nf)$ corrections in the flow equation for $g_2$.
However, the Thirring coupling $g_5^\star$ vanishes at any fixed point also beyond the large-$\Nf$ limit, as a consequence of the absence of a $g_5^2$ term also for finite $\Nf$.
For large, but finite, $\Nf$, in addition to the Gaussian fixed point at vanishing couplings, there are therefore $2^5 - 1 = 31$ possibly degenerate and/or complex fixed points at finite couplings.
In the large-$\Nf$ limit, the critical fixed points are located on the coordinate axes of our flavor-singlet basis.
The symmetry breaking pattern corresponding to each of the critical fixed points can then be simply identified on the basis of the mean-field decoupling of the corresponding flavor-singlet four-fermion term, which is controlled in the large-$\Nf$ limit.
This implies that the fixed point located at
\begin{equation}
    \boldsymbol{g}^\star_\text{GNH} = [1/4 + \mathcal O(1/\Nf),\mathcal O(1/\Nf),0,0,0,\mathcal O(1/\Nf^2)]\epsilon
\end{equation}
corresponds to Gross-Neveu-Heisenberg quantum criticality, with SU(2) order parameter $\langle \bar\psi^\alpha(\mathbb 1_2 \otimes \vec \sigma) \psi^\alpha\rangle$, and the fixed point located at
\begin{equation}
    \boldsymbol{g}^\star_\text{GNI} = [0,0,0,1/4+\mathcal O(1/\Nf),0,0]\epsilon
\end{equation}
corresponds to Gross-Neveu-Ising quantum criticality, with $\mathbbm Z_2$ order parameter $\langle \bar\psi^\alpha(\mathbb 1_2 \otimes \mathbb 1_2) \psi^\alpha\rangle$. Another fixed point located at
\begin{equation}
    \boldsymbol{g}^\star_{\text{GNI}'} = [0,0,0,0,0,1/4+\mathcal O(1/\Nf)]\epsilon
\end{equation}
also corresponds to Gross-Neveu-Ising quantum criticality, with $\mathbbm Z_2$ order parameter $\langle \bar\psi^\alpha(\gamma_5 \otimes \mathbb 1_2) \psi^\alpha\rangle$, breaking not only chiral symmetry but also inversion symmetry.
This fixed point is completely equivalent to the one at $\boldsymbol{g}_\text{GNI}^\star$ due to the invariance of the flow equations under exchange of the couplings as $(g_1, g_2, g_3, g_4, g_5, g_6) \leftrightarrow (-g_3, g_2, -g_1, -g_6, g_5, -g_4)$.
In particular, the set of eigenvalues of the stability matrix at $\boldsymbol{g}_\text{GNI}^\star$ and $\boldsymbol{g}_{\text{GNI}'}^\star$ are equal. This statement holds for arbitrary $\Nf$ within the one-loop approximation.

Note that the only flow equation containing a term $\propto g_4^2$ is the one for $g_4$ itself, such that upon starting the flow on the axis $\boldsymbol g \propto \boldsymbol g^\star_\text{GNI}$ in the ultraviolet, no other couplings $g_{i \neq 4}$ are generated in the infrared, in agreement with earlier work~\cite{bondi90, gracey16}.
Similarly, the flow equations for $g_3$, $g_4$, and $g_5$ do not contain terms $\propto g_i g_j$ with $i,j \in \{1,2,6\}$, such that $g_3$, $g_4$, and $g_5$ are not generated under the RG if simultaneously absent initially.
This is again a statement that holds for arbitrary $\Nf$ within the one-loop approximation.
Put differently, the space spanned by the couplings $g_1$, $g_2$, and $g_6$ is invariant under the one-loop RG.
As this subspace of the full theory space contains the Gross-Neveu-Heisenberg fixed point, we denote it as ``Gross-Neveu-Heisenberg subspace''.
For the analysis of the Gross-Neveu-Heisenberg criticality at the one-loop order, it is hence sufficient to consider the flow only within this subspace.
We emphasize, however, that the RG invariance of the Gross-Neveu-Heisenberg subspace is \emph{not} symmetry protected. At higher loop orders, there may be other terms that will require one to consider the full six-dimensional theory space.

%%%%%%%%%%%%%%%%%%%%%%%%%%%%%%%%%%%%%%%%%%%%%%%%%%%%%%%%%%%%%%%%%%%%%%%
\begin{table*}[tbh]
\caption{Fixed points in Gross-Neveu-Heisenberg subspace, their locations, number of relevant directions, and collisions as function of $\Nf$.}
\begin{tabular*}{\linewidth}{@{\extracolsep{\fill}} lllp{20.5em}}
\hline \hline
\multicolumn{1}{c}{} & \multicolumn{1}{l}{$(g_1^\star,g_2^\star,g_6^\star)/\epsilon$} & \multicolumn{1}{l}{$\#(\Theta_i > 0)$} & \multicolumn{1}{c}{} \\ \hline
$\mathcal{O}$  &  $(0,0,0)$  &  0  &  Gaussian\\
$\mathcal{H}$  &  $(h_1,h_2,h_6)$  &  1  & Gross-Neveu-Heisenberg\\
$\mathcal{I}$  &  $(0,0,-\frac{\Nf}{4\Nf-2})$  &  
$\begin{cases}
      1,& \text{for }\Nf > \Nf^{(1)} \\
      2,& \text{for }\Nf < \Nf^{(1)} \\
\end{cases}$
&  
\begin{tabular}{@{}l@{}}
\multicolumn{1}{@{}p{20.5em}@{}}{
Gross-Neveu-Ising$ ' $, collides with $\mathcal D$ for $\Nf \to \Nf^{(1)} = {3}/{2}$
}\end{tabular} \\
$\mathcal{A}$  &  $(0,-\frac{\Nf}{8},0)$  & 2  & part of RG invariant plane spanned by $\mathcal A$, $\mathcal B$, and $\mathcal C$\\
$\mathcal{B}$  &  $(\frac{\Nf}{4\Nf+4},0,-\frac{\Nf}{4\Nf+4})$  &  2  & part of RG invariant plane spanned by $\mathcal A$, $\mathcal B$, and $\mathcal C$  \\
$\mathcal{C}$  &  $(\frac{\Nf}{4\Nf+4},-\frac{\Nf}{8},-\frac{\Nf}{4\Nf+4})$  &  3  &  part of RG invariant plane spanned by $\mathcal A$, $\mathcal B$, and $\mathcal C$  \\
$\mathcal{D}$  &  $(e_1,e_2,e_6)$  & 
$\begin{cases}
      \text{complex},& \text{for }\Nf > \Nf^{(2)}  \\
      2, & \text{for }\Nf^{(1)} < \Nf < \Nf^{(2)} \\
      1, & \text{for }\Nf < \Nf^{(1)} \\
\end{cases}$
&  
\begin{tabular}{@{}l@{}}
\multicolumn{1}{@{}p{20.5em}@{}}{
annihilates with $\mathcal {E}$ for $\Nf \nearrow \Nf^{(2)} = 1.5146\dots$,
collides with $\mathcal {I}$ for $\Nf \to \Nf^{(1)} = {3}/{2}$
}\end{tabular} \\
$\mathcal{E}$  &  $(f_1,f_2,f_6)$  &
$\begin{cases}
      \text{complex},& \text{for }\Nf > \Nf^{(2)}  \\
      1, & \text{for }\Nf < \Nf^{(2)} \\
\end{cases}$ 
&   annihilates with $\mathcal {D}$ for $\Nf \nearrow \Nf^{(2)} = 1.5146\dots$ \\
\hline\hline
\end{tabular*}
\label{tab:HeisenbergSubspace}
\end{table*}
%%%%%%%%%%%%%%%%%%%%%%%%%%%%%%%%%%%%%%%%%%%%%%%%%%%%%%%%%%%%%%%%%%%%%%%

%%%%%%%%%%%%%%%%%%%%%%%%%%%%%%%%%%%%%%%%%%%%%%%%%%%%%%%%%%%%%%%%%%%%%%%
\subsection{Gross-Neveu-Heisenberg subspace}
\label{subsec:GNH-subspace}
%%%%%%%%%%%%%%%%%%%%%%%%%%%%%%%%%%%%%%%%%%%%%%%%%%%%%%%%%%%%%%%%%%%%%%%

\begin{figure}[tb]
    \includegraphics[width=\linewidth]{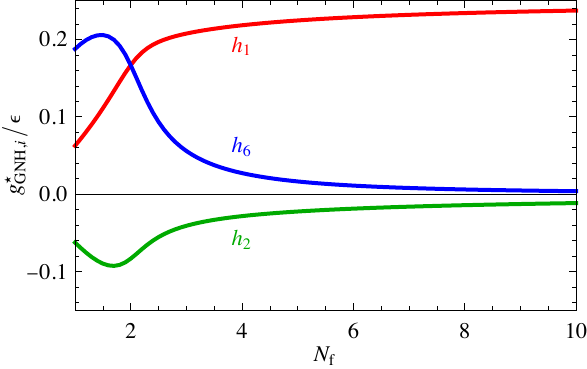}
    \caption{Evolution of Gross-Neveu-Heisenberg fixed-point couplings $\boldsymbol{g}^\star_\text{GNH} = [h_1(\Nf), h_2(\Nf), 0,0,0, h_6(\Nf)]\epsilon$ as function of flavor number $\Nf$.}
    \label{fig:GNHCouplings}
\end{figure}

\begin{figure*}[tbh]
    \centering
    \subfloat[$\Nf=2$]{\includegraphics[width=0.32\linewidth]{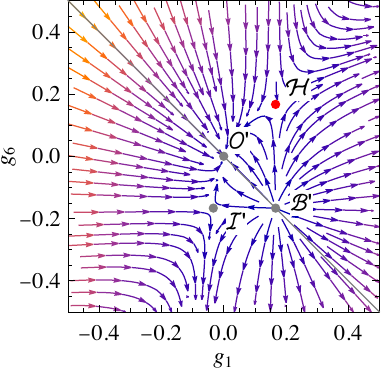}}\hfill
    \subfloat[$\Nf=4$]{\includegraphics[width=0.32\linewidth]{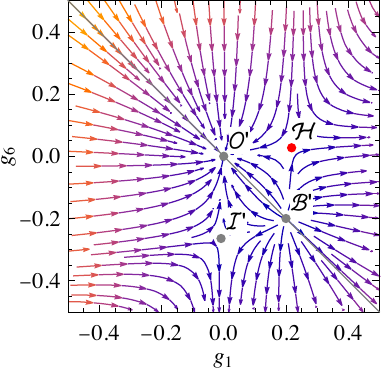}}\hfill
    \subfloat[$\Nf=8$]{\includegraphics[width=0.32\linewidth]{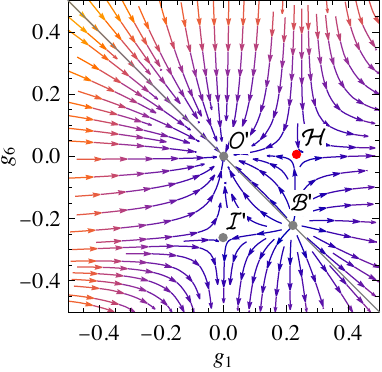}}
    \caption{RG flow in the plane spanned by $g_1$ and $g_6$ for fixed $g_2 = h_2(\Nf)$ through the Gross-Neveu-Heisenberg fixed point for (a)~$\Nf=2$, (b)~$\Nf=4$, and (c)~$\Nf=8$. 
    The red dot denotes the position of the Gross-Neveu-Heisenberg fixed point $\mathcal H$. 
    Gray dots labeled by $\mathcal O'$, $\mathcal B'$, and $\mathcal I'$ indicate points in parameter space in which the flow is perpendicular to the plane $g_2 = h_2(\Nf)$, which become projections of the Gaussian fixed point $\mathcal O$, the critical fixed point $\mathcal B$, and the bicritical fixed point $\mathcal I$ in the large-$\Nf$ limit.
    %Gray dots labeled by $\mathcal O'$, $\mathcal B'$, and $\mathcal I'$ indicate projections of the Gaussian fixed point $\mathcal O$, the critical fixed point $\mathcal B$, and the bicritical fixed point $\mathcal I$ onto the plane $g_2 = h_2(\Nf)$. 
    The gray line indicates the projection of the RG invariant subspace spanned by the fixed points $\mathcal A$, $\mathcal B$, and $\mathcal C$.
    }
    \label{fig:HeisenbergSubspace}
\end{figure*}

The Gross-Neveu-Heisenberg subspace is defined by
\begin{equation}
    \boldsymbol{g} =(g_1,g_2,0,0,0,g_6),
\end{equation}
which is the smallest RG invariant subspace containing the Gross-Neveu-Heisenberg fixed point. The latter is located at $\boldsymbol{g}_\text{GNH}^\star = [h_1(\Nf), h_2(\Nf), 0, 0, 0, h_6(\Nf)]\epsilon$, with real functions $h_1(\Nf) >0$, $h_2(\Nf) < 0$, and $h_6(\Nf) > 0$.
The Gross-Neveu-Heisenberg subspace also contains the Gross-Neveu-Ising$'$ fixed point at $\boldsymbol{g}_{\text{GNI}'}^\star = [0, 0, 0, 0, 0, -\Nf/(4\Nf-2)] \epsilon$, as well as six additional real or complex fixed points.
Table~\ref{tab:HeisenbergSubspace} shows the locations of all fixed points in the Gross-Neveu-Heisenberg subspace, together with their numbers of infrared relevant directions. 
The latter are obtained from the eigenvalues $\Theta_i$ of the stability matrix $(-\partial\beta_i/\partial g_j)$ at the respective fixed point.
Importantly, for all values of $\Nf$, the Gross-Neveu-Heisenberg fixed point features a single infrared relevant direction, corresponding to a critical fixed point.
For finite $\Nf < \infty$, it features finite fixed-point couplings in all three channels $- g_1/(2\Nf) [\bar\psi (\mathbb 1 \otimes \vec \sigma) \psi]^2$, $-g_2/(2\Nf) [\bar\psi (\gamma_\mu \otimes \vec \sigma) \psi]^2$, and $- g_6/(2\Nf) [\bar\psi (\gamma_5 \otimes \mathbb 1_2) \psi]^2$ of the Gross-Neveu-Heisenberg subspace.
The evolution of these fixed-point couplings as function of $\Nf$ is depicted in Fig.~\ref{fig:GNHCouplings}.
%
%%%%%%%%%%%%%%%%%%%%%%%%%%%%%%%%%%%%%%%%%%%%%%%%%%%%%%%%%%%%%%%%%%%%%%%
%%%%%%%%%%%%%%%%%%%%%%%%%%%%%%%%%%%%%%%%%%%%%%%%%%%%%%%%%%%%%%%%%%%%%%%
%
We have explicitly verified that perturbations out of the Gross-Neveu-Heisenberg subspace are infrared irrelevant in the vicinity of the Gross-Neveu-Heisenberg fixed point.
As an illustration of the RG flow near the Gross-Neveu-Heisenberg fixed point, Fig.~\ref{fig:HeisenbergSubspace} presents the flow diagram within the plane spanned by the couplings $g_1$ and $g_6$ for fixed $g_2 \equiv h_2(\Nf)$, for different values of $\Nf$. 
Therein, the Gross-Neveu-Heisenberg fixed point labeled by $\mathcal H$ is marked as red dot.
The gray dots labeled by $\mathcal O'$, $\mathcal B'$, and $\mathcal I'$ indicate points in parameter space in which the flow is perpendicular to the plane $g_2 \equiv h_2(\Nf)$. In the large-$\Nf$ limit, they represent projections of the fixed points $\mathcal O$, $\mathcal B$, and $\mathcal I$, respectively, and are adiabatically connected to these upon lowering $\Nf$.

%%%%%%%%%%%%%%%%%%%%%%%%%%%%%%%%%%%%%%%%%%%%%%%%%%%%%%%%%%%%%%%%%%%%%%%
%%%%%%%%%%%%%%%%%%%%%%%%%%%%%%%%%%%%%%%%%%%%%%%%%%%%%%%%%%%%%%%%%%%%%%%

The Gross-Neveu-Ising$ ' $ fixed point at $\boldsymbol{g}_{\text{GNI}'}^\star$ features a single relevant direction only for $\Nf$ above a critical flavor number $\Nf^{(1)} = 3/2 + \mathcal O(\epsilon)$. For $\Nf \to \Nf^{(1)}$, it collides with the bicritical fixed point $\mathcal D$, and exchanges, for $\Nf < \Nf^{(1)}$, its role with respect to RG stability with the latter.
Note that due to the symmetry of the flow equations, a simultaneous fixed-point collision occurs away from the Gross-Neveu-Heisenberg subspace at the same flavor number, involving the Gross-Neveu-Ising fixed point $\boldsymbol{g}_\text{GNI}^\star = [0, 0, 0, \Nf/(4\Nf-2), 0, 0]\epsilon$ and a bicritical fixed point at $[0,e_2,-e_1,-e_6,0,0]\epsilon$.
Such fixed-point-collision scenario, involving an exchange of stability between Gross-Neveu-Ising and bicritical fixed points, has been found previously in a one-loop analysis in fixed $D=2+1$ space-time dimensions~\cite{gehring15}.

Remarkably, for a second critical flavor number $\Nf^{(2)}= 1.5146\dots$, i.e., only slightly above $\Nf^{(1)}$, the fixed point $\mathcal D$ is involved in another fixed-point collision.
In this case, it merges with the fixed point $\mathcal E$, with both of them disappearing into the complex coupling plane for $\Nf > \Nf^{(2)}$.
Such fixed-point annihilation has been observed in a variety of gauge theories in 2+1~\cite{halperin74, nahum15, ihrig19, gies06, kaplan09, braun14, janssen16, herbut16b, gukov17} and higher~\cite{herbut14, janssen17} dimensions, but also in non-gauge theories~\cite{gehring15, herbut16a, gracey18c, gorbenko18, ma19, ma20, nahum20, weber22, hu22}.

%%%%%%%%%%%%%%%%%%%%%%%%%%%%%%%%%%%%%%%%%%%%%%%%%%%%%%%%%%%%%%%%%%%%%%%
\section{Critical exponents}
\label{sec:exponents}
%%%%%%%%%%%%%%%%%%%%%%%%%%%%%%%%%%%%%%%%%%%%%%%%%%%%%%%%%%%%%%%%%%%%%%%

The universal critical exponents we determine here are the correlation-length exponent $1/\nu$, the anomalous dimensions of the order-parameter and fermion fields, $\eta_\phi$ and $\eta_\psi$, respectively, and the corrections-to-scaling exponent $\omega$.
The dynamical critical exponent is $z=1$ exactly, as a consequence of Lorentz invariance, which has been shown to emerge at low energy for a Gross-Neveu-Heisenberg quantum critical point~\cite{ray21b}.

%%%%%%%%%%%%%%%%%%%%%%%%%%%%%%%%%%%%%%%%%%%%%%%%%%%%%%%%%%%%%%%%%%%%%%%
\subsection{Correlation-length exponent \(\boldsymbol{1/\nu}\)}
%%%%%%%%%%%%%%%%%%%%%%%%%%%%%%%%%%%%%%%%%%%%%%%%%%%%%%%%%%%%%%%%%%%%%%%

The correlation-length exponent $1/\nu$ determines the divergence of the correlation length near the quantum critical point. It is given by the unique positive eigenvalue $\Theta_1 > 0$ of the stability matrix $(-\partial\beta_i/\partial g_j)$ at the corresponding critical fixed point.
To leading order in the perturbative expansion, we find both for the Gross-Neveu-Heisenberg and the Gross-Neveu Ising fixed points
\begin{equation}\label{eq:nuinvGNH}
   1/\nu = \epsilon + \mathcal O(\epsilon),
\end{equation}
in agreement with the general result valid for all critical four-fermion models near the lower critical dimension~\cite{gehring15}.

%%%%%%%%%%%%%%%%%%%%%%%%%%%%%%%%%%%%%%%%%%%%%%%%%%%%%%%%%%%%%%%%%%%%%%%
\subsection{Order-parameter anomalous dimension \(\boldsymbol{\eta\sb\phi}\)}
%%%%%%%%%%%%%%%%%%%%%%%%%%%%%%%%%%%%%%%%%%%%%%%%%%%%%%%%%%%%%%%%%%%%%%%

As there appears no dangerously irrelevant coupling in the problem, we assume hyperscaling to hold.
The order-parameter anomalous dimension $\eta_\phi$ is then linked to the correlation-length exponent $1/\nu$ and the susceptibility exponent $\gamma$ via the hyperscaling relation
\begin{equation} \label{eq:hyperscaling}
    \eta_\phi = 2 - \gamma / \nu.
\end{equation}
Within our fermionic formulation, we can determine the susceptibility exponent $\gamma$ using the scheme described in Ref.~\cite{janssen16}.
To this end, we add the corresponding infinitesimal mass term to the effective Lagrangian as
\begin{align}
    \mathcal L \mapsto \mathcal L 
    + \Delta\,\bar\psi^\alpha \mathcal M \psi^\alpha,
\end{align}
with $\mathcal M = \mathbb 1_2 \otimes \mathbb 1_2$ for Gross-Neveu-Ising criticality and $\mathcal M = \mathbb 1_2 \otimes \vec \sigma$ for Gross-Neveu-Heisenberg criticality.
In the presence of the infinitesimal mass term, the scaling form of the free energy density near criticality reads~\cite{janssen16}
\begin{equation}
    f(\delta \boldsymbol{g}, \Delta) = |\delta \boldsymbol{g}|^{D \nu} \mathcal F^\pm\left(\frac{\Delta}{|\delta\boldsymbol{g}|^{x \nu}}\right),
\end{equation}
with scaling function $\mathcal F^\pm$. In the above equation, $\delta\boldsymbol{g}$ is the eigenvector associated with the RG relevant direction, and $x$ denotes the eigenvalue associated with the RG flow of $\Delta$,
\begin{equation}
    \beta_\Delta = - x \Delta + \mathcal O(\Delta^2).
\end{equation}
Differentiating twice with respect to the mass parameter $\Delta$ yields the scaling of the susceptibility,
\begin{equation}
    \chi = - \frac{\partial^2 f}{\partial \Delta^2} \propto |\delta \boldsymbol{g}|^{-\gamma},
    \quad
    \text{with}
    \quad
    \gamma = (2x - D) \nu.
\end{equation}
With the help of the hyperscaling relation~\eqref{eq:hyperscaling}, the order-parameter anomalous dimension is then given by
\begin{equation}
    \eta_\phi = D + 2(1 - x).
\end{equation}

At one-loop order, the flow of the mass parameter $\Delta$ has the form
\begin{equation}
    \beta_\Delta = - \left(1 + \sum_{i} c_i g_i \right)\Delta + \mathcal O(\Delta^2),
\end{equation}
with coefficients
\begin{align}
    c_i & = \frac{1}{4 D \Nf} \sum_{\mu} \Bigl\{
    \Nf \Tr (\mathcal M \gamma_\mu \mathcal O_i \gamma_\mu)
    \Tr (\mathcal M \mathcal O_i)
    \nonumber\\&\quad
    - \Tr\left[\mathcal O_i \gamma_\mu \mathcal M \gamma_\mu \mathcal O_i \mathcal M \right]
    \Bigr\},
    \label{eq:coefficients}
\end{align}
where $\mathcal O_i$ denotes the $4\times 4$ matrix in the four-fermion term parametrized by $g_i$, and for brevity we have omitted factors of $\mathbb 1_2$ in direct products with $\gamma$ matrices, i.e., $\gamma_\mu \equiv \gamma_\mu \otimes \mathbb 1_2$.
Note that in the above equation, no summation over repeated indices $i$ is assumed, and we have rescaled the couplings in the same way as described below Eqs.~\eqref{eq:betafunctions-1}--\eqref{eq:betafunctions-6}.
The order-parameter anomalous dimension at a fixed point $\boldsymbol{g}^\star = (g_i^\star)$ can then be obtained from
\begin{equation} \label{eq:etaphi-general}
    \eta_\phi = 2+\epsilon - 2\sum_i c_i g_i^\star
\end{equation}
in $D=2+\epsilon$ dimensions.

Evaluating the matrix algebra for the Gross-Neveu-Ising mass $\mathcal M = \mathbb 1_2 \otimes \mathbb 1_2$  and $\mathcal O_4 = \mathbb 1_2 \otimes \mathbb 1_2$, using the Gross-Neveu-Ising fixed-point value $\boldsymbol{g}_\text{GNI}^\star = [0, 0, 0, \Nf/(4\Nf - 2), 0, 0]\epsilon$, yields the order-parameter anomalous dimension for Gross-Neveu-Ising criticality
\begin{equation}
    \eta_\phi^\text{GNI} = 2 - \frac{2\Nf}{2\Nf - 1} \epsilon + \mathcal O(\epsilon^2).
\end{equation}
This agrees with the known results near the lower critical dimension~\cite{gracey90, gracey91, gracey08, gracey16}, thereby providing a first cross-check of our calculations.%
\footnote{Note that the definition for $N$ used in Ref.~\cite{gracey16} deviates from our definition for $\Nf$ as $N^\text{(Ref.~\cite{gracey16})} = 2\Nf^\text{(this work)}$.}

For Gross-Neveu-Heisenberg criticality, we assume $\mathcal M = \mathbb 1_2 \otimes \sigma_z$ without loss of generality, and use the couplings $\boldsymbol{g}_\text{GNH}^\star = [h_1(\Nf), h_2(\Nf), 0, 0, 0, h_6(\Nf)]\epsilon$ at the Gross-Neveu-Heisenberg fixed point.
Evaluating the matrix algebra yields
\begin{align} \label{eq:etaphiGNH-generalNf}
    \eta_\phi^\text{GNH} & = 2 + \left[1 - 2 \frac{\left(1 + 4\Nf\right) h_1(\Nf)+2h_2(\Nf)-h_6(\Nf)}{\Nf}\right]\epsilon 
    \nonumber\\&\quad
    + \mathcal{O}(\epsilon^2)
\end{align}
for general $\Nf$. 
It is instructive to further expand our small-$\epsilon$ results for large $\Nf$,
\begin{equation}
    \eta_\phi^\text{GNH} = 2 - \left[1 - \frac{1}{2\Nf} - \frac{5}{4\Nf^2} + \frac{19}{8\Nf^3} + \mathcal O(1/\Nf^4) \right] \epsilon + \mathcal O(\epsilon^2),
\end{equation}
which agrees, up to the order calculated, with the large-$\Nf$ exponents computed for arbitrary $2<D<4$~\cite{gracey18a}, upon expanding the latter for small $\epsilon = D-2$.%
\footnote{Note that the definitions for $\epsilon$ and $N$ used in Ref.~\cite{gracey18a} deviate from our definitions for $\epsilon$ and $\Nf$ as $\epsilon^\text{(Ref.~\cite{gracey18a})} = - \epsilon^\text{(this work)}/2$ and $N^\text{(Ref.~\cite{gracey18a})} = 2\Nf^\text{(this work)}$.}
This furnishes another nontrivial cross-check of our calculations.
For the cases relevant for interacting electrons on the single-layer~\cite{herbut06, herbut09, assaad14, otsuka16} and bilayer~\cite{ray21b, pujari16, ray18} honeycomb lattices, we explicitly find from Eq.~\eqref{eq:etaphiGNH-generalNf}, i.e., without expanding in $1/\Nf$,
\begin{align} \label{eq:etaphiGNH-Nf248}
    \eta_\phi^\text{GNH} = 
    \begin{cases} 
    2 - \frac{\epsilon}{6} + \mathcal O(\epsilon^2), & \text{for $\Nf = 2$}, \\
    2 - 0.812333\epsilon + \mathcal O(\epsilon^2), & \text{for $\Nf = 4$}, \\
    2 - 0.921305\epsilon + \mathcal O(\epsilon^2), & \text{for $\Nf = 8$}.
    \end{cases}
\end{align}
Equation~\eqref{eq:etaphiGNH-Nf248} represents one of the main results of this work.

%%%%%%%%%%%%%%%%%%%%%%%%%%%%%%%%%%%%%%%%%%%%%%%%%%%%%%%%%%%%%%%%%%%%%%%
\subsection{Fermion anomalous dimension \(\boldsymbol{\eta\sb\psi}\)}
%%%%%%%%%%%%%%%%%%%%%%%%%%%%%%%%%%%%%%%%%%%%%%%%%%%%%%%%%%%%%%%%%%%%%%%

%%%%%%%%%%%%%%%%%%%%%%%%%%%%%%%%%%%%%%%%%%%%%%%%%%%%%%%%%%%%%%%%%%%%%%%
\begin{figure*}[t]
    \subfloat[Gross-Neveu-Ising]{\includegraphics[width=0.48\linewidth]{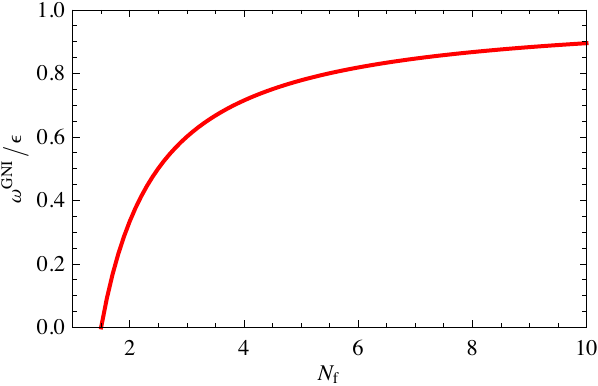}}\hfill
    \subfloat[Gross-Neveu-Heisenberg]{\includegraphics[width=0.48\linewidth]{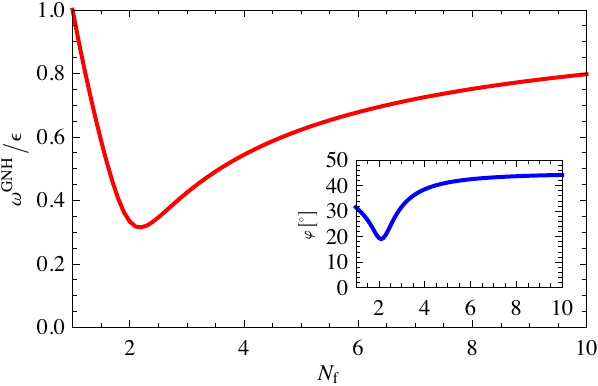}}
    \caption{%
    (a) Corrections-to-scaling exponent $\omega$ for the Gross-Neveu-Ising fixed point as a function of $\Nf$.
    (b)~Same as (a), but for the Gross-Neveu-Heisenberg fixed point, featuring a distinct minimum near $\Nf = 2$, corresponding to a slow flow towards the fixed point. The inset shows the angle $\varphi$ between $\boldsymbol{g}^\star_\text{GNH}$ and the surface normal $\boldsymbol{n} = (1, 0, 1)/\sqrt{2}$ of the invariant subspace spanned by the fixed points $\mathcal A$, $\mathcal B$, and $\mathcal C$, as function of $\Nf$.}
    \label{fig:omega}
\end{figure*}
%%%%%%%%%%%%%%%%%%%%%%%%%%%%%%%%%%%%%%%%%%%%%%%%%%%%%%%%%%%%%%%%%%%%%%%

As in any critical four-fermion model near the lower critical dimension, the fermion anomalous dimension vanishes at one-loop order,  $\eta_\psi = 0 + \mathcal O(\epsilon^2)$.
This implies that knowledge of the $\mathcal O(\epsilon)$ fixed-point values, together with the result of the corresponding two-loop selfenergy diagram, is sufficient to compute $\eta_\psi$ to order $\mathcal O(\epsilon^2)$.
While at the one-loop order all regulator dependences can be factored out by appropriate rescalings of the couplings, this may no longer be true at higher orders.
We employ a minimal subtraction scheme analogous to Ref.~\cite{bondi90}, with an infrared cutoff in the form of a mass term $m \bar\psi (\mathbb 1_2 \otimes \mathbb 1_2) \psi$, and an effective fermion propagator
\begin{equation}
    G(p) = \frac{-\rmi p_\mu (\gamma_\mu \otimes \mathbb 1_2)}{p^2 + m^2}.
\end{equation}
Note that we have omitted the mass term in the numerator of the effective propagator, which gives no contribution to the pole in $1/\epsilon$, as a consequence of the infrared finiteness of the theory~\cite{bondi90}.
Evaluating the sunset diagram for the fermion selfenergy at a fixed point $\boldsymbol{g}^\star = (g_i^\star)$ yields
\begin{align} \label{eq:etapsi-general}
    \eta_\psi = g_i^{\star} H_{ij} g_j^\star,
\end{align}
with the matrix elements
\begin{align}
    H_{ij} & = 
   \dfrac{1}{32\Nf^2}\sum_{\mu,\nu,\lambda} (\delta_{\mu,0} \delta_{\nu\lambda} + \delta_{\nu,0} \delta_{\mu \lambda} + \delta_{\lambda,0} \delta_{\mu\nu})
    \bigl\{
    \nonumber \\ & \quad
    \Nf \Tr(\gamma_0 \mathcal O_i \gamma_\mu \mathcal O_j) \Tr(\gamma_\nu \mathcal O_j \gamma_\lambda \mathcal O_i) 
    \nonumber \\ & \quad
    - \Tr(\gamma_0 \mathcal{O}_i  \gamma_\mu \mathcal{O}_j \gamma_\nu \mathcal{O}_i \gamma_\lambda \mathcal{O}_j)
    \bigr\},
    \label{eq:matrixelements}
\end{align}
where $\mathcal O_i$ again denotes the $4\times 4$ matrix in the four-fermion term parametrized by $g_i$, and for brevity we have omitted factors of $\mathbb 1_2$ in direct products with $\gamma$ matrices, i.e., $\gamma_\mu \equiv \gamma_\mu \otimes \mathbb 1_2$.
Note that in the above equation, no summation over repeated indices $i$ and $j$ is assumed, and we have rescaled the couplings as $g_i/2\pi \mapsto g_i$, 
which agrees with the rescaling below Eqs.~\eqref{eq:betafunctions-1}--\eqref{eq:betafunctions-6} for the present regularization scheme.

Evaluating the matrix algebra for the Gross-Neveu-Ising fixed point $\boldsymbol{g}_\text{GNI}^\star$ yields
\begin{equation}
    \eta_\psi^\text{GNI} = \frac{4\Nf - 1}{8(2\Nf-1)^2} \epsilon^2 + \mathcal O(\epsilon^3),
\end{equation}
in agreement with the literature results~\cite{gracey90, gracey91, gracey08, gracey16}, providing another cross-check of our approach.

For the Gross-Neveu-Heisenberg fixed point $\boldsymbol{g}_\text{GNH}^\star = [h_1(\Nf), h_2(\Nf), 0, 0, 0, h_6(\Nf)]\epsilon$, we find
\begin{align}
    \eta_\psi^\text{GNH} & = \Bigl[
    3(4\Nf + 1)h_1^2 + 24 \Nf h_2^2  + (4\Nf-1) h_6^2
    \nonumber\\&\quad
    + 12 h_1 h_2 - 6 h_1 h_6 + 12 h_2 h_6
    \Bigr]\frac{\epsilon^2}{2\Nf^2}+\mathcal{O}(\epsilon^3)
\end{align}
for general $\Nf$, leading to
\begin{equation} \label{eq:etapsiGNH-largeN}
    \eta_\psi^\text{GNH} = \left[\frac{3}{2\Nf} - \frac{9}{8\Nf^2} - \frac{3}{4\Nf^3} + \mathcal O(1/\Nf^4) \right]\epsilon^2 + \mathcal O(\epsilon^3)
\end{equation}
in the large-$\Nf$ limit.
The first two terms agree with the previous large-$\Nf$ calculation in fixed space-time dimension $2<D<4$~\cite{gracey18a}, when expanding the latter for small $\epsilon = D-2$. The third term $\propto 1/\Nf^3$ does not agree: We have found $-{3\epsilon^2}/(4\Nf^3)$, whereas Ref.~\cite{gracey18a} suggests $-9\epsilon^2/(8\Nf^3)$.
However, this discrepancy can be traced back to a term $-2/[3(\mu-1)]$ on the right-hand side of Eq.~(6.6) of Ref.~\cite{gracey18a}, which should not be there. Without that term, the large-$\Nf$ result, when expanded near two dimensions, fully agrees with our Eq.~\eqref{eq:etapsiGNH-largeN}.\footnote{We are grateful to John Gracey for pointing this out to us.}
For the physically relevant cases~\cite{herbut06, herbut09, assaad14, otsuka16, ray21b, pujari16, ray18}, we find
\begin{equation}
    \eta_\psi^\text{GNH} = 
    \begin{cases}
    \frac{7}{72}\epsilon^2 + \mathcal O(\epsilon^3), & \text{for } \Nf = 2, \\
    0.0748866\epsilon^2 + \mathcal O(\epsilon^3), & \text{for } \Nf = 4, \\
    0.0422519\epsilon^2 + \mathcal O(\epsilon^3), & \text{for } \Nf = 8,
    \end{cases}
\end{equation}
which represents another important result of our work.

%%%%%%%%%%%%%%%%%%%%%%%%%%%%%%%%%%%%%%%%%%%%%%%%%%%%%%%%%%%%%%%%%%%%%%%
\subsection{Corrections-to-scaling exponent \(\boldsymbol{\omega}\)}
\label{subsec:omega}
%%%%%%%%%%%%%%%%%%%%%%%%%%%%%%%%%%%%%%%%%%%%%%%%%%%%%%%%%%%%%%%%%%%%%%%

The exponent $\omega$ determines the leading corrections to scaling near the quantum critical point. It is given by the negative of the second-largest eigenvalue $\Theta_2 < 0$ of the stability matrix $(-\partial\beta_i/\partial g_j)$ at the corresponding critical fixed point.
The leading-order results for the Gross-Neveu-Ising and Gross-Neveu-Heisenberg fixed points are shown in Figs.~\ref{fig:omega}(a) and~(b), respectively.

In the Gross-Neveu-Ising case, the corrections-to-scaling exponent $\omega$ vanishes for $\Nf \to \Nf^{(1)} = 3/2$. This is a direct consequence of the fixed-point collision occurring at this value of $\Nf$. For $\Nf < \Nf^{(1)}$, the Gross-Neveu-Ising fixed point develops a second relevant direction, as discussed in Sec.~\ref{subsec:GNH-subspace}.

Remarkably, in the Gross-Neveu-Heisenberg case, the exponent features a distinct minimum of $\omega \approx 0.3 \epsilon + \mathcal O(\epsilon^2)$ near $\Nf = 2$.
This can be understood to arise from the competition between the different interaction channels in the vicinity of the critical fixed point. In particular, the infrared relevant direction of the Gross-Neveu-Heisenberg fixed point, which is parallel to the fixed-point vector $\boldsymbol{g}_\text{GNH}^\star$ itself, has a large component perpendicular to the RG invariant plane spanned by the fixed points $\mathcal A$, $\mathcal B$, and $\mathcal C$. 
This RG invariant plane is characterized by an O(4) symmetry generated by $(\mathbbm 1_2, \gamma_5) \otimes \vec \sigma$, under which the bilinears $\bar\psi(\mathbbm 1_2 \otimes \vec \sigma) \psi$ and $\rmi\bar\psi(\gamma_5 \otimes \mathbbm 1_2)\psi$ transform as components of an O(4) vector, and which is an enhancement of the spin SU(2) symmetry defined in Eq.~\eqref{eq:spinSUtwo}.
This is illustrated in the inset of Fig.~\ref{fig:omega}(b), which shows the angle $\varphi$ between $\boldsymbol{g}_\text{GNH}^\star$ and the surface normal $\boldsymbol n = (1,0,1)/\sqrt{2}$ of the RG invariant plane spanned by the fixed points $\mathcal A$, $\mathcal B$, and $\mathcal C$, featuring a distinct minimum near $\Nf = 2$. The presence of an RG invariant plane in the vicinity of the Gross-Neveu-Heisenberg fixed point perpendicular to the fixed point's relevant direction arguably leads to a slow flow on the critical surface, i.e., towards the critical point. Near $\Nf=2$, the corrections-to-scaling exponent $\omega$ is therefore relatively small, implying that fluctuations over a comparatively large number of length scales need to be integrated out to approach the ultimate infrared behavior.
We emphasize that this result arises from the competition between the different interaction channels within our Fierz-complete basis, and could not have been obtained within standard $4-\varepsilon$ or large-$\Nf$ approaches, which typically involve the fluctuations within the respective condensation channel only.

%%%%%%%%%%%%%%%%%%%%%%%%%%%%%%%%%%%%%%%%%%%%%%%%%%%%%%%%%%%%%%%%%%%%%%%
\section{Estimates for Gross-Neveu-Heisenberg criticality in 2+1 dimensions}
\label{sec:estimates}
%%%%%%%%%%%%%%%%%%%%%%%%%%%%%%%%%%%%%%%%%%%%%%%%%%%%%%%%%%%%%%%%%%%%%%%

The knowledge of the critical exponents in $D=2+\epsilon$ space-time dimensions, together with literature results for these exponents in $D= 4- \varepsilon$ dimensions~\cite{zerf17}, allows us to employ an interpolational resummation scheme in order to obtain estimates for the critical exponents in the physical cases for $D = 2+1$. In the Gross-Neveu-Ising case, such an approach has previously been shown to lead to a significant improvement~\cite{janssen14, ihrig18}.
Here, we focus on the leading exponents $1/\nu$, $\eta_\phi$, and $\eta_\psi$. We do not attempt an interpolation of the corrections-to-scaling exponent $\omega$, since the leading corrections to scaling, as obtained in the previous subsection, arise from the competition between different interaction channels, which has not been included in $4-\varepsilon$ expansion approaches to date.

We employ a scheme based on two-sided Pad\'e approximants defined as
\begin{equation}\label{eq:PadeApproximant}
    [m/n](D):=\frac{a_0 + a_1 D + \dots + a_m D^m}{1 + b_1 D + \dots + b_n D^n},
\end{equation}
with non-negative integers $m$ and $n$, and real coefficients $a_0, \dots, a_m$ and $b_1, \dots, b_n$, chosen such that the Pad\'e approximant matches both the $2+\epsilon$ and $4-\epsilon$ results, when expanding the approximant near the lower and upper critical dimensions, respectively.
The order $m+n$ of the Pad\'e approximant is determined by the number of constraints given by the $2+\epsilon$ and $4-\varepsilon$ results.
Near the upper critical dimension, all leading exponents are known up to quartic order in $\varepsilon = 4-D$.
Near the lower critical dimension, we have computed the correlation-length exponent $1/\nu$ and the boson anomalous dimension $\eta_\phi$ to linear order, and the fermion anomalous dimension to quadratic order in $\epsilon = D-2$.
This implies that the orders of the corresponding Pad\'e approximants are $m+n = 6$ for $1/\nu$ and $\eta_\phi$, and $m+n = 7$ for $\eta_\psi$, respectively.
While in principle several choices for $m$ and $n$ are possible, some of these cannot satisfy all constraints near $D=2$ and $D=4$ for real coefficients. This applies to $m=0$ for $\eta_\phi$ and $1/\nu$, as well as to $m=0,1,2$ for $\eta_\psi$.
Furthermore, some choices lead to singularities of the corresponding Pad\'e approximants between $2<D<4$.

%%%%%%%%%%%%%%%%%%%%%%%%%%%%%%%%%%%%%%%%%%%%%%%%%%%%%%%%%%%%%%%%%%%%%%%
\begin{table}[tbp!]
\caption{Critical exponents of the Gross-Neveu-Heisenberg universality class for $\Nf=2$ four-component Dirac fermions in $D=3$ space-time dimensions, relevant for the transition between the Dirac semimetal and the antiferromagnetic insulator in the Hubbard model on the honeycomb lattice~\cite{herbut09, assaad14, otsuka16}.
Here, we have used different two-sided Pad\'e approximants $[m/n]$, interpolating between the expansions near the lower and upper critical dimensions.
In the upper (lower) part of the table, marked as $\mathcal O(\epsilon, \varepsilon^4)$ [$\mathcal O(\epsilon^2, \varepsilon^4)$], we have employed the results to linear (quadratic) order in $\epsilon=D-2$ and to quartic order in $\varepsilon = 4-D$. The latter are obtained from Ref.~\cite{zerf17}.
Approximants that cannot satisfy all constraints are marked as ``n.e.'', those exhibiting singularities in $2<D<4$ dimensions are marked as ``sing.''
The dashes ``$-$'' signify approximants for which the required $\epsilon^2$ corrections are not yet available.}
\begin{tabular*}{\linewidth}{@{\extracolsep{\fill}} lcccc} 
\noalign{\smallskip}\hline\hline\noalign{\smallskip}
$\Nf=2$ & $[m/n]$ & $1/\nu$ & $\eta_\phi$ & $\eta_\psi$\\
\noalign{\smallskip}\hline\noalign{\smallskip}
$\mathcal O(\epsilon, \varepsilon^4)$ &  $[1/5]$ & 0.83569 & 1.07386 &  \text{n.e.}\\
  &  $[2/4]$  &  0.76888  &  1.02731  & \text{n.e.}\\
  &  $[3/3]$  &  \text{sing.} & 0.97641  & \text{sing.}\\
  &  $[4/2]$  &  \text{sing.} & 0.95000  & 0.15943\\
  &  $[5/1]$  &  0.75902 & 1.02035 & \text{sing.}\\
  &  $[6/0]$  &  0.94485 & 1.03755  & 0.15592\\
  \noalign{\smallskip}
$\mathcal O(\epsilon^2, \varepsilon^4)$ &  $[3/4]$ & $-$ & $-$ & 0.14750\\
  &  $[4/3]$ & $-$ & $-$ & 0.10199\\
  &  $[5/2]$ & $-$ & $-$ & \text{sing.}\\
  &  $[6/1]$ & $-$ & $-$ & \text{sing.}\\
  &  $[7/0]$ & $-$ & $-$ & 0.13960\\
\noalign{\smallskip}\hline\hline
\end{tabular*}
\label{tab:CritExpNf2}
\end{table}
%%%%%%%%%%%%%%%%%%%%%%%%%%%%%%%%%%%%%%%%%%%%%%%%%%%%%%%%%%%%%%%%%%%%%%%

%%%%%%%%%%%%%%%%%%%%%%%%%%%%%%%%%%%%%%%%%%%%%%%%%%%%%%%%%%%%%%%%%%%%%%%
\begin{figure*}[tbp!]
\includegraphics[width=.33\linewidth]{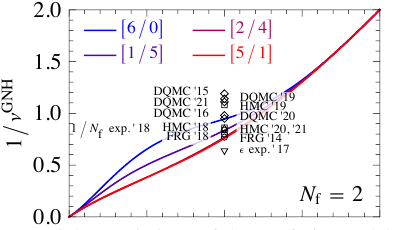}\hfill
\includegraphics[width=.33\linewidth]{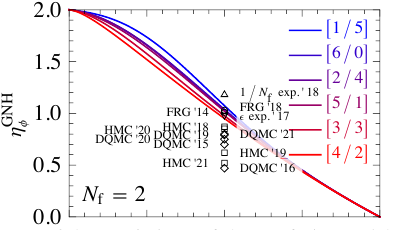}\hfill
\includegraphics[width=.34\linewidth]{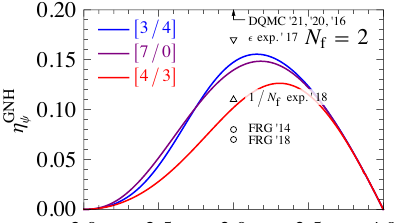}\\
\includegraphics[width=.33\linewidth]{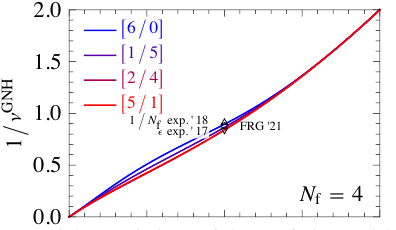}\hfill
\includegraphics[width=.33\linewidth]{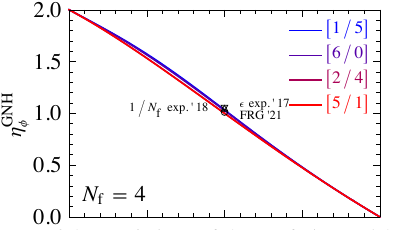}\hfill
\includegraphics[width=.34\linewidth]{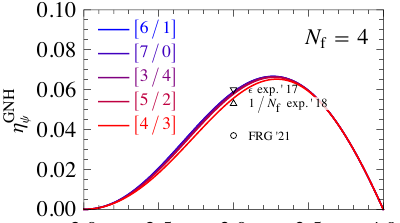}\\
\includegraphics[width=.33\linewidth]{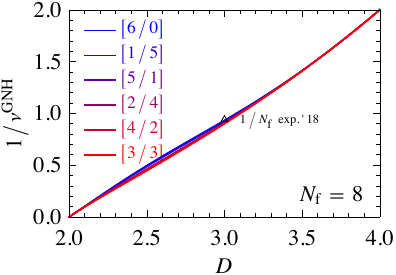}\hfill
\includegraphics[width=.33\linewidth]{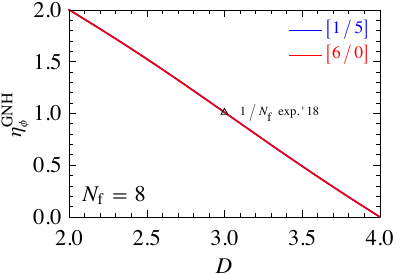}\hfill
\includegraphics[width=.34\linewidth]{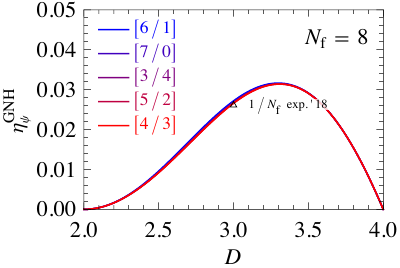}
\caption{Correlation-length exponent $1/\nu$ (left column), order-parameter anomalous dimension $\eta_\phi$ (center column), and fermion anomalous dimension $\eta_\psi$ (right column) of the Gross-Neveu-Heisenberg universality class as function of space-time dimension $2<D<4$ for $\Nf = 2$ (first row), $\Nf = 4$ (second row), and $\Nf=8$ (last row). The different curves in each panel correspond to different Pad\'e approximants $[m/n]$, which interpolate between the series expansions around the lower and upper critical dimensions $D=2$ and $D=4$, respectively. 
Data points at $D=3$ refer to literature results from $4-\varepsilon$ expansion~($\triangledown$)~\cite{zerf17}, $1/\Nf$ expansion~($\triangle$)~\cite{gracey18a, ray21b, ray18}, functional RG~($\circ$)~\cite{janssen14, knorr18, ray21b}, as well as determinantal quantum Monte Carlo~($\diamond$)~\cite{toldin15, liu19, liu21, otsuka16, otsuka20} and hybrid Monte Carlo ($\square$)~\cite{buividovich18, buividovich19, ostmeyer20, ostmeyer21} simulations.
%Black dots at $D=3$ refer to literature results from $4-\varepsilon$ expansion~\cite{zerf17}, $1/\Nf$ expansion~\cite{gracey18a, ray21b, ray18}, functional RG~\cite{janssen14, knorr18, ray21b}, as well as determinantal quantum Monte Carlo (DQMC)~\cite{toldin15, liu19, liu21, otsuka16, otsuka20} and hybrid Monte Carlo (HMC)~\cite{buividovich18, buividovich19, ostmeyer20, ostmeyer21} simulations.
}
\label{fig:CritExpInterpolation}
\end{figure*}
%%%%%%%%%%%%%%%%%%%%%%%%%%%%%%%%%%%%%%%%%%%%%%%%%%%%%%%%%%%%%%%%%%%%%%%

Figure~\ref{fig:CritExpInterpolation} shows the non-singular Pad\'e approximants for the critical exponents at the Gross-Neveu-Heisenberg fixed point as a function of space-time dimension $2<D<4$ for the physically relevant cases on the single-layer~\cite{herbut06, herbut09, assaad14, otsuka16} and bilayer~\cite{ray21b, pujari16, ray18} honeycomb lattices, i.e., for $\Nf=2$, $\Nf=4$, and $\Nf=8$.
The numerical values in $D=2+1$ space-time dimensions of the different Pad\'e approximants are given in Tables~\ref{tab:CritExpNf2}, \ref{tab:CritExpNf4}, and \ref{tab:CritExpNf8}, respectively.
%
%%%%%%%%%%%%%%%%%%%%%%%%%%%%%%%%%%%%%%%%%%%%%%%%%%%%%%%%%%%%%%%%%%%%%%%
\begin{table}[tbp!]
\caption{Same as Table \ref{tab:CritExpNf2}, but for $\Nf=4$ four-component Dirac fermions, relevant for the transition between nematic and coexistent nematic-antiferromagnetic orders on the Bernal-stacked honeycomb bilayer~\cite{ray21b}.}
\begin{tabular*}{\linewidth}{@{\extracolsep{\fill}} lcccc} 
\noalign{\smallskip}\hline\hline\noalign{\smallskip}
$\Nf=4$ & $[m/n]$ & $1/\nu$ & $\eta_\phi$ & $\eta_\psi$\\
\noalign{\smallskip}\hline\noalign{\smallskip}
$\mathcal O(\epsilon, \varepsilon^4)$ & $[1/5]$ & 0.86441 & 1.03391  &  \text{sing.}\\
& $[2/4]$  & 0.84006 & 1.00147  & \text{sing.}\\
& $[3/3]$  & \text{sing.} & \text{sing.} & 0.06418\\
& $[4/2]$  & \text{sing.} & \text{sing.} & 0.05950\\
& $[5/1]$  &  0.83956 & 0.99998  & \text{sing.}\\
& $[6/0]$  &  0.89489 & 1.02706  & 0.05906\\
\noalign{\smallskip}
$\mathcal O(\epsilon^2, \varepsilon^4)$ & $[3/4]$ & $-$ & $-$ & 0.05848\\ 
& $[4/3]$  & $-$ & $-$ & 0.05570\\
& $[5/2]$  & $-$ & $-$ & 0.05776\\ 
& $[6/1]$  & $-$ & $-$ & 0.05887\\
& $[7/0]$  & $-$ & $-$ & 0.05886\\
\noalign{\smallskip}\hline\hline
\end{tabular*}
\label{tab:CritExpNf4}
\end{table}
%%%%%%%%%%%%%%%%%%%%%%%%%%%%%%%%%%%%%%%%%%%%%%%%%%%%%%%%%%%%%%%%%%%%%%%
%
%%%%%%%%%%%%%%%%%%%%%%%%%%%%%%%%%%%%%%%%%%%%%%%%%%%%%%%%%%%%%%%%%%%%%%%
\begin{table}[tbp!]
\caption{Same as Table \ref{tab:CritExpNf2}, but for $\Nf=8$ four-component Dirac fermions, relevant for the transition between the trigonal-warping-induced Dirac semimetal and the antiferromagnetic insulator on the Bernal-stacked honeycomb bilayer~\cite{pujari16, ray18}.}
\begin{tabular*}{\linewidth}{@{\extracolsep{\fill}} lcccc} 
\noalign{\smallskip}\hline\hline\noalign{\smallskip}
$\Nf=8$ & $[m/n]$ & $1/\nu$ & $\eta_\phi$ & $\eta_\psi$\\
\noalign{\smallskip}\hline\noalign{\smallskip}
$\mathcal O(\epsilon, \varepsilon^4)$ & $[1/5]$ & 0.92445  &  1.01040  & \text{n.e.} \\
& $[2/4]$  & 0.90752 & \text{sing.} & \text{n.e.}\\
& $[3/3]$  & 0.89871 & \text{sing.} & 0.02769 \\
& $[4/2]$  & 0.90389 & \text{sing.} & 0.02679 \\
& $[5/1]$  & 0.90865 & \text{sing.} & \text{sing.}\\
& $[6/0]$  & 0.93249 & 1.00979 & 0.02591\\
\noalign{\smallskip}
$\mathcal O(\epsilon^2, \varepsilon^4)$ & $[3/4]$  & $-$ & $-$ & 0.02682\\ 
& $[4/3]$  & $-$ & $-$ & 0.02657\\
& $[5/2]$  & $-$ & $-$ & 0.02660\\ 
& $[6/1]$  & $-$ & $-$ & 0.02718\\
& $[7/0]$  & $-$ & $-$ & 0.02676\\
\noalign{\smallskip}\hline\hline
\end{tabular*}
\label{tab:CritExpNf8}
\end{table}
%%%%%%%%%%%%%%%%%%%%%%%%%%%%%%%%%%%%%%%%%%%%%%%%%%%%%%%%%%%%%%%%%%%%%%%
%
The final best-guess estimates are determined via averaging over the results from the different non-singular Pad\'e approximants for the highest-order expansion results available for each exponent.
We thus arrive at
\begin{equation}
1/\nu = 
\begin{cases}
0.83(12), & \text{for } \Nf = 2,\\
0.86(4), & \text{for } \Nf = 4,\\
0.913(20), & \text{for } \Nf = 8,
\end{cases}
\end{equation}
for the correlation-length exponent, as well as
\begin{equation}
\eta_\phi =
\begin{cases}
1.014(64), & \text{for } \Nf = 2,\\
1.016(18), & \text{for } \Nf = 4,\\
1.0101(3), & \text{for } \Nf = 8,
\end{cases}
\end{equation}
and
\begin{equation}
\eta_\psi =
\begin{cases}
0.130(28), & \text{for } \Nf = 2,\\
0.0579(22), & \text{for } \Nf = 4,\\
0.0268(4), & \text{for } \Nf = 8,
\end{cases}
\end{equation}
for the boson and fermion anomalous dimensions, respectively.
In the above equations, the numbers in parentheses correspond to the maximal deviations from the mean values among the different approximants, which can be understood as a lower bound for the uncertainty of our best-guess estimates. We emphasize that the true systematic error is hard to quantify and may be significantly larger than this lower bound. This is particularly true for cases in which only few non-singular Pad\'e approximants exist. Nevertheless, we find it reassuring that for the case of $\Nf = 8$, for which the large-$\Nf$ expansion is expected to yield reliable results, our estimates for $1/\nu$ and $\eta_\phi$ are within error bars fully consistent with the large-$\Nf$ results quoted in Ref.~\cite{ray18}, and our estimate for $\eta_\psi$ is within error bars almost consistent with those of Ref.~\cite{ray18}.%
\footnote{Note that the definition for $N$ used in Ref.~\cite{ray18} deviates from our definition for $\Nf$ as $2N^\text{(Ref.~\cite{ray18})} =  \Nf^\text{(this work)}$.}

Our estimates are compared with a variety of literature results available for $\Nf=2$ from $4-\varepsilon$ expansion~\cite{zerf17}, $1/\Nf$ expansion~\cite{gracey18a}, functional RG~\cite{janssen14, knorr18}, as well as determinantal quantum Monte Carlo~\cite{toldin15, liu19, liu21, otsuka16, otsuka20, xu21} and hybrid Monte Carlo~\cite{buividovich18, buividovich19, ostmeyer20, ostmeyer21} simulations in Table~\ref{tab:literature}.
Available literature results for different $\Nf$ are included as black dots in Fig.~\ref{fig:CritExpInterpolation}.
For $\Nf=2$, the deviations between the results of the different methods are considerable: For $1/\nu$, the analytical estimates are typically significantly smaller than those of determinantal quantum Monte Carlo calculations; the hybrid Monte Carlo estimates lie roughly between these two. For $\eta_\phi$, on the other hand, the analytical estimates are significantly larger than those of most of the quantum Monte Carlo simulations. In the case of $\eta_\psi$, analytical estimates are again significantly smaller than those of determinantal quantum Monte Carlo calculations; hybrid Monte Carlo estimates for $\eta_\psi$ are not available at present.
For $\Nf=4$, literature results from $4-\varepsilon$ expansion~\cite{zerf17}, $1/\Nf$ expansion~\cite{gracey18a}, and functional RG~\cite{ray21b}, all of which as compiled in Ref.~\cite{ray21b}, agree very well with our results in the case of $1/\nu$ and $\eta_\phi$; some deviations, in particular from the functional RG estimate, are present in the case of $\eta_\psi$.

%%%%%%%%%%%%%%%%%%%%%%%%%%%%%%%%%%%%%%%%%%%%%%%%%%%%%%%%%%%%%%%%%%%%%%%
\begin{table}[tbp]
\caption{Gross-Neveu-Heisenberg critical exponents for $\Nf=2$ four-component Dirac fermions from interpolation between series expansions near lower and upper critical dimensions (this work) in comparison with previous results from fourth-order $4-\varepsilon$ expansion~\cite{zerf17}, second-order (third-order for $\eta_\psi$) $1/\Nf$ expansion~\cite{gracey18a}, functional RG in local potential approximation (LPA')~\cite{janssen14} and next-to-leading order derivative expansion (NLO)~\cite{knorr18}, as well as determinantal quantum Monte Carlo (DQMC)~\cite{toldin15, liu19, liu21, otsuka16, otsuka20, xu21} and hybrid Monte Carlo (HMC)~\cite{buividovich18, buividovich19, ostmeyer20, ostmeyer21} simulations for different linear lattice sizes $L$ and inverse temperatures $\beta$ or projection times $\tau$.
In cases where $\eta_\phi$ and/or $1/\nu$ were not computed directly, we have employed appropriate hyperscaling relations to obtain these.
In cases where results from different Pad\'e approximants ($4-\varepsilon$ expansion), different regulators (functional RG), or different lattices (Monte Carlo simulations) are available within the same work, we show the corresponding mean values.}
\begin{tabular*}{\linewidth}{@{\extracolsep{\fill}} lclll} 
\noalign{\smallskip}\hline\hline\noalign{\smallskip}
$\Nf=2$ & Year & \multicolumn{1}{c}{$1/\nu$} & \multicolumn{1}{c}{$\eta_\phi$} & \multicolumn{1}{c}{$\eta_\psi$} \\
\noalign{\smallskip}\hline\noalign{\smallskip}
Interpolation (this work) & 2022 & 0.83(12) & 1.01(6) & 0.13(3)\\
$4-\varepsilon$ expansion, $\mathcal O(\varepsilon^4)$~\cite{zerf17} & 2017 & 0.64 & 0.98  & 0.17 \\
$1/\Nf$ expansion, $\mathcal O(1/\Nf^{2,3})$~\cite{gracey18a} & 2018 & 0.85 & 1.18 & 0.11 \\
functional RG, NLO \cite{knorr18} & 2018 & 0.80 & 1.03 & 0.07\\
functional RG, LPA' \cite{janssen14} & 2014 & 0.77 & 1.01 & 0.08\\
DQMC, $\tau \sim L \leq 40$ \cite{otsuka20} & 2020 & 0.95(5) & 0.75(4) & 0.23(4) \\
DQMC, $\tau \sim L \leq 40$ \cite{otsuka16} & 2016 & 0.98(1) & 0.47(7)  & 0.22(2)\\
DQMC, $\beta = L\leq 24$ \cite{liu21} & 2021 & 1.11(4) & 0.80(9) & 0.29(2) \\
DQMC, $\beta = L \leq 21$ \cite{liu19} & 2019 & 1.14(9) & 0.79(5) & -- \\
DQMC, $\tau = 60$, $L \leq 18$ \cite{toldin15} & 2015 & 1.19(6) & 0.70(15) &  -- \\
DQMC, $\beta = L \leq 20$ \cite{xu21} & 2021 & 1.01(8) & 0.55(2) & -- \\
HMC, $\beta \leq 12$, $L \leq 102$ \cite{ostmeyer21} & 2021 & 0.84(4) & 0.52(1) & -- \\
HMC, $\beta \leq 12$, $L \leq 102$ \cite{ostmeyer20} & 2020 & 0.84(4) & 0.85(13) & -- \\
HMC, $\beta = 21, L \leq 24$ \cite{buividovich19} & 2019 & 1.08 & 0.62 &  -- \\
HMC, $\beta = 21, L \leq 18$ \cite{buividovich18} & 2018 & 0.86 & 0.87(2) &  -- \\
\noalign{\smallskip}\hline\hline
\end{tabular*}
\label{tab:literature}
\end{table}
%%%%%%%%%%%%%%%%%%%%%%%%%%%%%%%%%%%%%%%%%%%%%%%%%%%%%%%%%%%%%%%%%%%%%%%

%%%%%%%%%%%%%%%%%%%%%%%%%%%%%%%%%%%%%%%%%%%%%%%%%%%%%%%%%%%%%%%%%%%%%%%
\section{Conclusions}
\label{sec:conclusions}
%%%%%%%%%%%%%%%%%%%%%%%%%%%%%%%%%%%%%%%%%%%%%%%%%%%%%%%%%%%%%%%%%%%%%%%

To conclude, we have determined the critical behavior of the Gross-Neveu-Heisenberg universality class within an $\epsilon$ expansion around the lower critical space-time dimension of two.
In contrast to the Gross-Neveu-Ising case~\cite{gracey16}, the critical fixed point associated with the Gross-Neveu-Heisenberg universality class is characterized by a combination of different four-fermion interaction channels, requiring an approach that takes these channels into account in an unbiased way.
For the Gross-Neveu-Heisenberg case, a Fierz-complete basis of the theory space compatible with the symmetries of the model comprises six four-fermion interaction terms.
Applying the general formula derived in Ref.~\cite{gehring15} to this system has allowed us to derive the flow equations of this six-dimensional theory space.
By making use of hyperscaling relations and the flow of an infinitesimal symmetry breaking fermion bilinear, we have demonstrated how to compute the full set of critical exponents within the fermionic language.
Applying this scheme to the Gross-Neveu-Ising fixed point, for which various literature results are available, facilitates a nontrivial cross-check of our approach. Our results for the leading-order order-parameter anomalous dimension $\eta_\phi$ and the next-to-leading order fermion anomalous dimension $\eta_\psi$ at the Gross-Neveu-Heisenberg fixed point are original.

These results have allowed us to obtain improved estimates for the critical exponents in $D=2+1$ space-time dimensions, as relevant for interacting fermion models on the honeycomb and bilayer honeycomb lattices. Here, we have employed a resummation scheme that takes the expansions near the lower and upper critical dimensions simultaneously into account. For the Gross-Neveu-Ising case, such an interpolational approach has previously been shown to provide significantly more reliable estimates in comparison with standard extrapolation schemes~\cite{janssen14, ihrig18}.
In the Gross-Neveu-Heisenberg case, our results for $\Nf=8$, relevant for the transition between the trigonal-warping-induced semimetal and the antiferromagnetic insulator on the Bernal-stacked honeycomb bilayer~\cite{pujari16, ray18}, agree with previous large-$\Nf$ estimates~\cite{gracey18a} within an uncertainty on the level of $\lesssim 3\%$.
For $\Nf=4$, relevant for the nematic-to-coexistence transition on the honeycomb bilayer~\cite{ray21b}, the deviations between our estimates and the large-$\Nf$ results~\cite{gracey18a}, upon appropriate resummation of the latter~\cite{ray21b}, are only slightly larger as compared with the $\Nf=8$ case, with the largest relative difference of $\simeq 8\%$ occurring for the fermion anomalous dimension.

Interestingly, for $\Nf=2$, relevant for the semimetal-to-antiferromagnet transition in the honeycomb-lattice Hubbard model~\cite{herbut09, assaad14, otsuka16}, we have found that the Gross-Neveu-Heisenberg fixed point is characterized by a slow flow towards criticality, corresponding to a small corresponding exponent $\omega$ and generically sizable corrections to scaling. This result can be understood to arise from the competition between different interaction channels present at the Gross-Neveu-Heisenberg fixed point.
This is in contrast to the Gross-Neveu-Ising fixed point, which is characterized by a single and uniquely identifiable interaction channel.
The critical point of a lattice model of, e.g., spinless fermions interacting via a repulsive nearest-neighbor density-density interaction, can therefore be close in theory space to the Gross-Neveu-Ising fixed point, leading to small scaling corrections.
The generically large scaling corrections in the Gross-Neveu-Heisenberg case for $\Nf=2$ might explain the significant spread between the estimates from the various numerical and analytical approaches, cf.\ Table~\ref{tab:literature}.
To track down the origin of these discrepancies, it would be interesting to test whether the data obtained in the simulations are in principle compatible with a small corrections-to-scaling exponent $\omega$.
Within our one-loop analysis, we estimate $\omega \approx 0.3$ for $\Nf=2$; however, a more accurate estimate, obtained from, e.g., a full two-loop analysis around the lower critical dimension, or an interpolation between the lower and upper critical dimensions, would certainly be highly desirable.
An interpolational approach to estimate $\omega$ would require to compute the scaling dimensions of the different four-fermion terms within the $4-\varepsilon$ expansion, which might be an interesting direction for future work.

On more general grounds, our work demonstrates how to determine the critical behavior of fermion models in cases where the corresponding critical fixed point is characterized by different four-fermion interaction channels.
This should be of relevance for other fermionic universality classes as well.
In particular, our general formulas for the order-parameter anomalous dimension $\eta_\phi$ to linear order in $\epsilon = D-2$, see Eqs.~\eqref{eq:coefficients} and \eqref{eq:etaphi-general}, and the fermion anomalous dimension $\eta_\psi$ to quadratic order in $\epsilon$, see Eqs.~\eqref{eq:etapsi-general} and \eqref{eq:matrixelements}, together with the general formula for the flow equations of relativistic four-fermion models~\cite{gehring15}, could be immediately applied to other relativistic universality classes, such as Gross-Neveu-XY~\cite{bobev15, li17, classen17}, Gross-Neveu-SO($N$)~\cite{seifert20, ray21a, janssen22}, or nematic~\cite{vojta00, schwab22} transitions.
These may host even more interesting phenomena, such as emergent supersymmetry~\cite{lee07, jian15, gies17}, fixed-point annihilation and complexification~\cite{gehring15, janssen22}, or quasiuniversal behavior~\cite{schwab22}.

%%%%%%%%%%%%%%%%%%%%%%%%%%%%%%%%%%%%%%%%%%%%%%%%%%%%%%%%%%%%%%%%%%%%%%%
\begin{acknowledgments}
%%%%%%%%%%%%%%%%%%%%%%%%%%%%%%%%%%%%%%%%%%%%%%%%%%%%%%%%%%%%%%%%%%%%%%%
%
We thank John Gracey for very valuable discussions and comments on the manuscript, and Michael Scherer for collaborations on related topics.
This work has been supported by the Deutsche Forschungsgemeinschaft (DFG) through SFB 1143 (A04 and A07, Project No.~247310070), the W\"{u}rzburg-Dresden Cluster of Excellence {\it ct.qmat} (EXC 2147, Project No.~390858490), and the Emmy Noether program (ME4844/1-1, Project No.~327807255, and JA2306/4-1, Project No.~411750675).
\end{acknowledgments}

%%%%%%%%%%%%%%%%%%%%%%%%%%%%%%%%%%%%%%%%%%%%%%%%%%%%%%%%%%%%%%%%%%%%%%%
% BIBLIOGRAPHY: FOR USE WITH BIBTEX
%%%%%%%%%%%%%%%%%%%%%%%%%%%%%%%%%%%%%%%%%%%%%%%%%%%%%%%%%%%%%%%%%%%%%%%
\bibliographystyle{longapsrev4-2}
\bibliography{GNH-eps}
%%%%%%%%%%%%%%%%%%%%%%%%%%%%%%%%%%%%%%%%%%%%%%%%%%%%%%%%%%%%%%%%%%%%%%%

\end{document}